# Ferroelectric topological superconductor

Xiaoming Zhang[1*], Pei Zhao[1], and Feng Liu[2*]

[1]*College of Physics and Optoelectronic Engineering, Ocean University of China, Qingdao, Shandong 266100, China*

[2] *Department of Materials Science and Engineering, University of Utah, Salt Lake City, Utah 84112, USA*

[*]Correspondence to: fliu@eng.utah.edu (F.L.) and zxm@ouc.edu.cn (X.Z.)

Two-dimensional topological superconductor (TSC) represents an exotic quantum material with quasiparticle excitation manifesting in dispersive Majorana mode (DMM) at the boundaries. A domain-wall DMM can arise at the boundary between two TSC domains with opposite Chern numbers or with a π-phase shift in their pairing gap, which can only be tuned by magnetic field. Here we propose the concept of ferroelectric (FE) TSC, which not only enriches the domain-wall DMMs but also significantly makes them electrically tunable. The π-phase shift of the pairing gap is shown to be attained between two TSC domains of opposite FE polarization, and switchable by reversing FE polarizations. In combination with ferromagnetic (FM) polarization, the domain wall can host helical, doubled chiral, and fused DMMs, which can be transferred into each other by changing the direction of electrical and/or magnetic field. Furthermore, based on first-principles calculations, we demonstrate $\alpha$-In$_2$Se$_3$ to be a promising FE TSC candidate in proximity with a FM layer and a superconductor substrate. We envision that FE TSC will significantly ease the manipulation of DMM by electrical field to realize fault-tolerant quantum computation.

## I. INTRODUCTION

Topological superconductor (TSC) represents an attractive condensed-matter system that can manifest an analogous form of long-sought Majorana fermion [1,2]. A two-dimensional (2D) TSC is featured with 1D dispersive Majorana modes (DMMs) at the boundaries, whose transport induced non-Abelian braiding statistics offers a promising route to fault-tolerant quantum computing [3-5]. Intrinsic TSC phase at the weak pairing limit of spin-triplet $p$-wave superconductivity (SC) is characterized by an odd-parity pairing (OPP) gap [6,7], but occurs rarely in nature. Alternatively, an effective OPP gap can arise extrinsically from constructive interplay between magnetic field, $s$-wave pairing gap, and spin-momentum-locked states, which have attracted much recent attention [8-14]. In addition to an outer boundary, the DMMs may arise at an inner boundary between two TSC domains with opposite Chern numbers or with a $\pi$-phase shift in their OPP gap, and can only be tuned by switching the ferromagnetic (FM) field. Here, we propose the concept of 2D ferroelectric (FE) TSC, rendering a new TSC platform that enables the domain-wall DMMs to be electrically tunable.

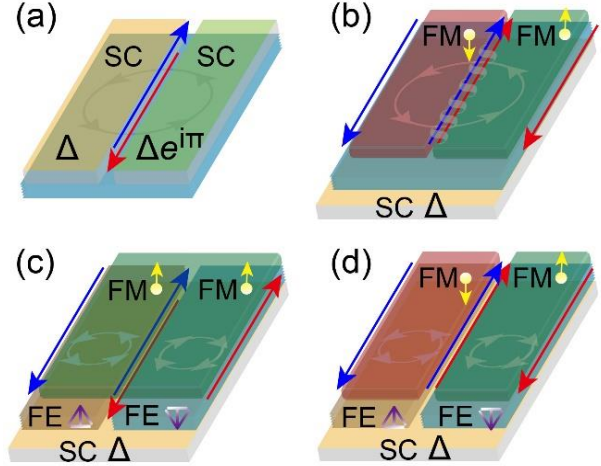

FIG. 1. Representative TSC platforms with tunable DMMs. (a) The helical DMMs realized by adding a $\pi$-phase shift to the pairing gap $\Delta$ of SC. (b) The fused and chiral DMMs realized by opposite FM fields under a uniform pairing gap. (c) The helical and chiral DMMs realized by opposite FE polarizations under a uniform pairing gap and FM field. (d) The doubled and chiral DMMs realized by opposite FE polarizations and opposite FE fields under a uniform pairing gap. The DMMs are plotted by red- and blue-colored lines with the arrow indicating the transport direction. The dot with yellow arrow shows the direction of FM field and the bar with purple arrow shows the direction of FE polarization. The gray circle with arrows represents the spin-momentum-locked states with specific chirality. The wrapped DMMs at the FM-FM domain wall in (b) denotes the fused DMMs.

DMMs exhibiting distinct transport characteristics can be engineered by magnetic field in various existing TSC platforms. The extrinsic TSC phase was initiated by placing superconductors on surface of 3D topological insulators (TIs) [8], which induces helical DMMs around the interface where an *s*-wave pairing gap receives a π-phase shift (Fig. 1a), i.e. the gaps having opposite sign. The π-phase shift has been achieved by some clever and delicate experimental designs, such as by controlling magnetic flux through a SC loop in a Josephson junction [15,16], engineering the width and magnetization in the SC-FM-SC junction [17,18], or by shifting exactly half unit-cell of FeSe$_{0.45}$Te$_{0.55}$ lattice [19,20]. The helical DMMs will evolve into chiral ones when time-reversal symmetry is broken by interfacing with a FM film, which leads to rich TSC phases tuned by magnetic field [9,10,21,22], and a fused domain-wall DMM could arise at the interface of two TSC domains of opposite Chern number or FM field (Fig. 1b). Evidence of chiral DMMs has been observed in magnetic TI (Cr$_{0.12}$Bi$_{0.26}$Sb$_{0.62}$)$_2$Te$_3$ grown on Nb substrate [23]. Another TSC scheme replaces the 3D TI with a 2D non-centrosymmetric materials exploiting Rashba spin-orbit coupling (SOC) effect [11,12], which is again only magnetically tunable as the Rashba SOC in a given non-centrosymmetric material is fixed. Since the Rashba SOC naturally arises at the interface, a simplified TSC platform for this scheme is to place a FM Shiba lattice on a SC substrate [13,14], whose chiral DMMs are experimentally confirmed in heterostructures, including Pb/Co/Si(111) [24], Fe/Re(0001)-O(2×1) [25], and CrBr$_3$/NbSe$_2$ [26], via proximity effect.

In this article, we demonstrate a unique class D TSC platform realized by a FE layer with the proximity effects of FM layer and SC substrate (Fig. 1c and 1d), so that the switchable FE polarization endows the FE TSC with a tunable Rashba SOC. Interestingly, we found that the chirality of SOC plays the same role as the π-phase shift of *s*-wave pairing in changing the sign of OPP gap, because the spin-momentum-locked states have also a sense of sign to distinguish their chiral spin textures. However, the spin texture is generally fixed for a given material or an existing TSC platform. In contrast, FE TSC enables the sign of OPP to be easily switched by reversing the chirality of Rashba SOC via FE polarization, namely by electrical field. Together with the sign of Chern number *C* controlled by FM polarization, there can be at least four different TSC domains having different sign combinations of OPP and *C*. Remarkably, depending on the FE/FM field configurations of adjacent domains, the domain-wall can host helical (Fig. 1c), doubled chiral (Fig. 1d), and fused DMMs (Fig. 1b), which can be transferred into each other by reversing electrical and/or magnetic field. Such high tunability may facilitate different braiding operations by engineering the location of domain walls and the DMM transport properties and even moving the domain-wall DMM by applying a dynamic field. Lastly, we confirm our theoretical model by first-principles calculations demonstrating an *α*-In$_2$Se$_3$-based FE TSC in proximity with a FM layer and a SC substrate.

## II. RESULTS

### A. Theory of FE TSC

We will recast the effective OPP theory originally based on Rashba SOC [11-14] in the framework of FE TSC by constructing a single-orbital tight-binding (TB) model on a square lattice (Fig. 2a). Under the basis of $\phi_{\mathbf{k}} = (c_{\mathbf{k}\uparrow}, c_{\mathbf{k}\downarrow})^T$, TB Hamiltonian is

$$h = \sum_{\mathbf{k}} \phi_{\mathbf{k}}^\dagger h(\mathbf{k}) \phi_{\mathbf{k}} \text{ with } h(\mathbf{k}) = \epsilon_{\mathbf{k}} + V_z \sigma_z + 2\lambda \mathbf{g}_{\mathbf{k}} \cdot \boldsymbol{\sigma}, \quad (1)$$

where $c_{\mathbf{k}\alpha}$ ($c_{\mathbf{k}\alpha}^\dagger$) annihilates (creates) an electron with spin $\alpha$ and momentum $\mathbf{k}$. $\mathbf{k} = (k_x, k_y)$, $\epsilon_{\mathbf{k}} = 2t(cosk_x + cosk_y) - \mu$, $\mathbf{g}_{\mathbf{k}} = (sink_y, -sink_x)$, $\boldsymbol{\sigma} = (\sigma_x, \sigma_y)$. $\sigma_i$ ($i=x, y, z$) denotes the Pauli matrices. $t$ and $\mu$ are respectively the nearest-neighbor electron hopping and chemical potential. $\lambda$ represents the strength of build-in FE field and $V_z$ the external FM field. The reversal of FE (FM) field is modeled by changing the sign of $\lambda$ ($V_z$). $\mathbf{g}_{\mathbf{k}} \cdot \boldsymbol{\sigma}$ describes the spin-momentum locking induced by Rashba SOC.

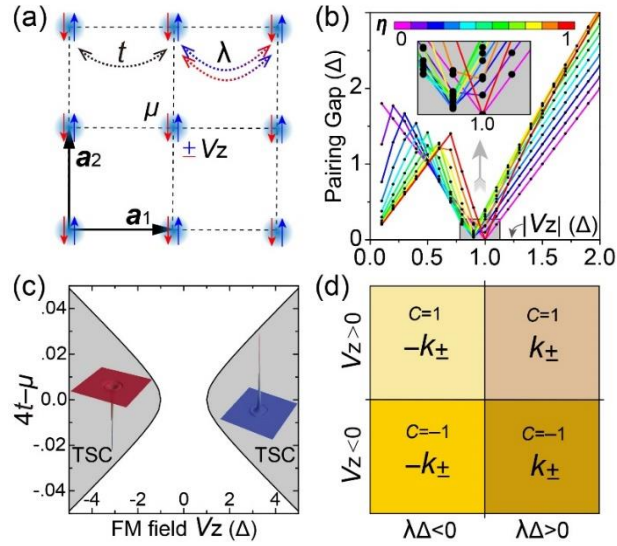

FIG. 2. FE TSC TB theory. (a) Schematics of the TB model on a square lattice. (b) FM field $V_z$ dependent pairing gap at $\mathbf{k} = 0$ calculated by diagonalizing $H_{BdG}^h(\mathbf{k})$ with different $\eta$. Inset is the magnified view near the FE TSC phase transition. (c) FE TSC phase diagram in the parameter space of $V_z$ and $4t - \mu$. $V_z$ is in the unit of $\Delta$, with $\Delta = 0.1t$. The left and right inset shows distribution of Berry curvature with $V_z < 0$ and $V_z > 0$, respectively. (d) Four FE TSC domains with different sign combinations of OPP and $C$.

Then the Bogoliubov-de-Gennes (BdG) Hamiltonian is constructed to describe the superconducting states under the Nambu basis of $\psi_{\mathbf{k}} = (c_{\mathbf{k}\uparrow}, c_{\mathbf{k}\downarrow}, c_{-\mathbf{k}\uparrow}^\dagger, c_{-\mathbf{k}\downarrow}^\dagger)^T$:

$$H_{BdG} = \frac{1}{2} \sum_k \psi_{\mathbf{k}}^\dagger H_{BdG}(\mathbf{k}) \psi_{\mathbf{k}}, \text{ with}$$

$$H_{BdG}(\mathbf{k}) = \begin{pmatrix} \epsilon_{\mathbf{k}} + V_z & 2i\lambda k_- & 0 & \Delta \\ -2i\lambda k_+ & \epsilon_{\mathbf{k}} - V_z & -\Delta & 0 \\ 0 & -\Delta & -\epsilon_{\mathbf{k}} - V_z & -2i\lambda k_+ \\ \Delta & 0 & 2i\lambda k_- & -\epsilon_{\mathbf{k}} + V_z \end{pmatrix}, \quad (2)$$

where $k_\pm = \sin k_x \pm i \sin k_y$. The s-wave pairing potential $\Delta$ takes positive real values. The $H_{\text{BdG}}(\mathbf{k})$ can be transformed into the partitioned matrices $H_{\text{BdG}}^{\text{t}}(\mathbf{k}) = T_l H_{\text{BdG}}(\mathbf{k}) T_r$ with

$$H_{\text{BdG}}^{\text{t}}(\mathbf{k}) = \frac{1}{\epsilon_{\mathbf{k}}-V_z}\begin{pmatrix} \xi_{\mathbf{k}} & 4i\lambda\Delta k_- & 0 & 0 \\ -4i\lambda\Delta k_+ & -\xi_{\mathbf{k}} & 0 & 0 \\ 0 & 0 & (\epsilon_{\mathbf{k}} - V_z)^2 & 0 \\ 0 & 0 & 0 & -(\epsilon_{\mathbf{k}} - V_z)^2 \end{pmatrix}, \quad (3)$$

$$T_l = \frac{1}{\epsilon_{\mathbf{k}}-V_z}\begin{pmatrix} \epsilon_{\mathbf{k}} - V_z & -2i\lambda k_- & 0 & \Delta \\ 0 & \Delta & \epsilon_{\mathbf{k}} - V_z & -2i\lambda k_+ \\ 0 & 0 & \epsilon_{\mathbf{k}} - V_z & 0 \\ 0 & 0 & 0 & \epsilon_{\mathbf{k}} - V_z \end{pmatrix}, \quad (4)$$

$$T_r = \frac{1}{\epsilon_{\mathbf{k}}-V_z}\begin{pmatrix} \epsilon_{\mathbf{k}} - V_z & 0 & 0 & 0 \\ 2i\lambda k_+ & \Delta & \epsilon_{\mathbf{k}} - V_z & 0 \\ 0 & \epsilon_{\mathbf{k}} - V_z & 0 & 0 \\ \Delta & 2i\lambda k_- & 0 & \epsilon_{\mathbf{k}} - V_z \end{pmatrix}. \quad (5)$$

Here $\xi_{\mathbf{k}} = \epsilon_{\mathbf{k}}^2 + \Delta^2 - V_z^2 - 4\lambda^2 k_+ k_-$ and the basis of $H_{\text{BdG}}^{\text{t}}(\mathbf{k})$ is $\psi_{\mathbf{k}}^t = \left(c_{\mathbf{k}\uparrow}, c_{-\mathbf{k}\uparrow}^\dagger, c_{\mathbf{k}\downarrow} - \frac{2i\lambda k_+ c_{\mathbf{k}\uparrow}+\Delta c_{-\mathbf{k}\uparrow}^\dagger}{\epsilon_{\mathbf{k}}-V_z}, c_{-\mathbf{k}\downarrow}^\dagger - \frac{2i\lambda k_- c_{-\mathbf{k}\uparrow}^\dagger+\Delta c_{\mathbf{k}\uparrow}}{\epsilon_{\mathbf{k}}-V_z}\right)^T$. One can easily realize that the submatrix $\begin{pmatrix} \xi_{\mathbf{k}} & 4i\lambda\Delta k_- \\ -4i\lambda\Delta k_+ & -\xi_{\mathbf{k}} \end{pmatrix}$ describes a p-wave TSC having the OPP $\Xi k_\pm$; here $\Xi = sign(\lambda\Delta)$ and the subscript "$\pm$" indicating a time-reversal pair of degenerate chiral states. A TSC phase emerges at the weak pairing limit for the effective OPP [6,7], i.e. $\xi_{\mathbf{k}} < 0$ at the center of Brillouin zone ($\mathbf{k} = 0$); it corresponds to the TSC phase transition condition of $(4t - \mu)^2 + \Delta^2 < V_z^2$.

Notice that the above transformation of BdG Hamiltonian is not a unitary transformation because $T_l \neq T_r^{-1}$; the eigenvalues of $H_{\text{BdG}}^{\text{t}}(\mathbf{k})$ and $H_{\text{BdG}}(\mathbf{k})$ are different. When $H_{\text{BdG}}(\mathbf{k})$ is gradually transformed into $H_{\text{BdG}}^{\text{t}}(\mathbf{k})$, the varying eigenvalues can be numerically traced by constructing a hypothetical Hamiltonian, $H_{\text{BdG}}^{\text{h}}(\mathbf{k}) = (1-\eta)H_{\text{BdG}}(\mathbf{k}) + \eta H_{\text{BdG}}^{\text{t}}(\mathbf{k})$. By diagonalizing $H_{\text{BdG}}^{\text{h}}(\mathbf{k})$, we calculate the pairing gap at $\mathbf{k} = 0$ as a function of FM field $V_z$ for different $\eta$ and setting $4t - \mu = 0$ (Fig. 2b). One can clearly see a gap reopening process with the increasing $V_z$, where the gap closes at $V_z = \Delta$ for both $\eta = 0.0$ and $\eta = 1.0$, but the gap never closes when $\eta$ is varied from 0.0 to 1.0 for a given $V_z$. This can be also captured from the different band dispersions of superconducting quasi-particles when $\eta$ varies from 0.0 to 1.0 for the trivial SC phase (Fig. S1a), at the critical point of phase transition (Fig. S1b), and for the FE TSC phase (Fig. S1c), respectively, in the Supplemental Material (SM) [27] that includes Ref. [28-33]. This indicates that the pairing gaps of $H_{\text{BdG}}(\mathbf{k})$ and $H_{\text{BdG}}^{\text{t}}(\mathbf{k})$ are adiabatically connected belonging to the same topological class. Thus, the FE TSC phase described by $H_{\text{BdG}}(\mathbf{k})$ will emerge under the condition $(4t - \mu)^2 + \Delta^2 < V_z^2$ (Fig. 2c). Given that $4t$ is the energy of the Kramers pair (KP) induced by Rashba SOC at $\mathbf{k} = 0$, the condition derived here for TSC phase transition is consistent with previous reports [11,12].

## B. DMMs of FE TSC tuned by electrical and magnetic field

A critical insight from the FE TSC theory is that the effective OPP gap possesses the form of $\Xi k_\pm$ with $\Xi = sign(\lambda\Delta)$, meaning the sign of OPP gap depends on the sign of $\lambda$, in addition to the phase of *s*-wave pairing gap. Thus, the OPP symmetry of $k_\pm$ and $-k_\pm$ can be interchanged by either attaching a $\pi$-phase to the pairing gap, which is well known before, or reversing the sign of $\lambda$, as we propose here. Together with different signs of *C* (Fig. 2c), there are four FE TSC domains that can be achieved and distinguished by different *C* and/or OPP (Fig. 2d). Here the Chern number *C* depends on the direction of FM field, i.e. $C=sign(V_z)$, as confirmed by integrating the Berry curvature of BdG quasi-particle wave function below the pairing gap (see insets of Fig. 2c).

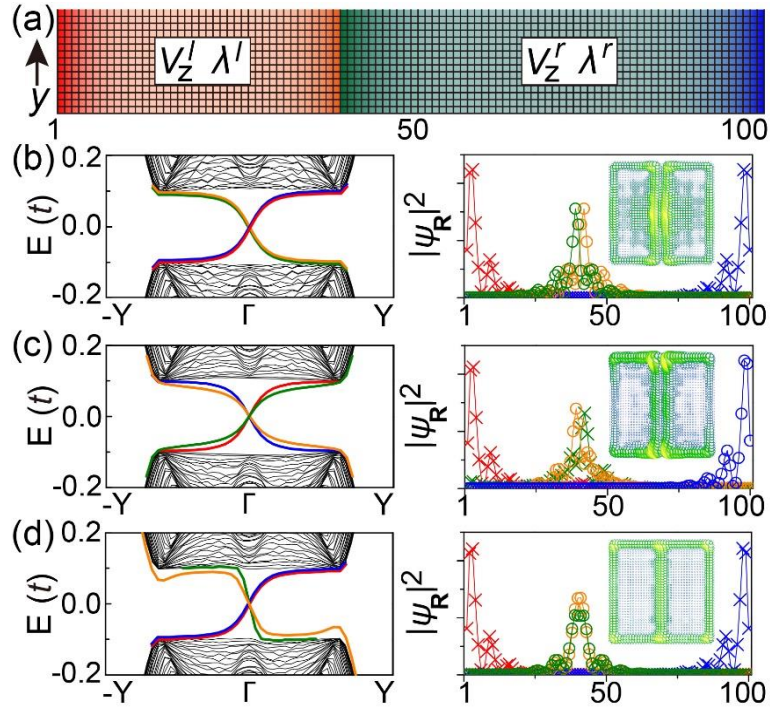

FIG. 3. DMMs with high tunability. (a) Schematics of a nanoribbon, with $(V_z^l, \lambda^l)$ and $(V_z^r, \lambda^r)$ marking the left and right domains, respectively. The width of left/right domain is 40/60 unit-cells and the arrow *y* indicates the periodic direction. (b-d) Nanoribbon BdG quasi-particle band structure (left panel) and real-space DMM distribution $|\psi_R|^2$ (right panel) calculated for the left/right domain configurations of (b) $\{V_{z+}^l\lambda_+^l\}/\{V_{z-}^r\lambda_-^r\}$ or $\{V_{z+}^l\lambda_-^l\}/\{V_{z-}^r\lambda_+^r\}$, (c) $\{V_{z+}^l\lambda_+^l\}/\{V_{z+}^r\lambda_-^r\}$ or $\{V_{z+}^l\lambda_-^l\}/\{V_{z+}^r\lambda_+^r\}$, and (d) $\{V_{z+}^l\lambda_+^l\}/\{V_{z-}^r\lambda_+^r\}$ or $\{V_{z+}^l\lambda_-^l\}/\{V_{z-}^r\lambda_-^r\}$, respectively. We use curly braces to show the combinations of $V_z$ and $\lambda$, whose sign is indicated by the subscript + or −. The DMMs inside the pairing gap (left panels) and the corresponding real-space distributions (right panels) are plotted in the same color. Crosses and circles indicate the DMMs transporting in the +*y*- and −*y*-direction, respectively. Insets are the distributions of the DMMs in a square island of FE TSC with a domain-wall in the middle.

In conventional TSC platforms, the sign of $\lambda$ is not changeable; while here for FE TSC, we generalize the TSC domains by including also the sign of $\lambda$ or OPP gap. The four FE TSC domains (Fig. 2d) will lead to $4^N$ configurations for a system with $N$ FE and/or FM domains, where the domain-wall boundaries interfacing different TSC domains host a rich variety of DMMs. The domain walls can be formed between FE domains of opposite polarization, FM domains of opposite magnetization, and their combinations, which we abbreviate as the FE-FE, FM-FM, and FE/FM-FE/FM domain wall, respectively. As an example, we consider a FE TSC nanoribbon made of two domains denoted by $(V_z^l, \lambda^l)$ and $(V_z^r, \lambda^r)$ on the left and right, respectively, as shown in Fig. 3a. By changing the signs of these four parameters, 16 configurations can be divided into four categories based on the chirality and location of DMMs, each consisting of four configurations.

The first category has the left and right domains of the same TSC phase, which requires $sign\ (V_z^l) = sign\ (V_z^r)$ and $sign\ (\lambda^l) = sign\ (\lambda^r)$. There is only the outer-edge DMMs whose transport direction can be reversed by changing the FM field direction (see Fig. S2 in SM [27]). The second category has the interface between two FE TSC domains with opposite sign of both $C$ and $\lambda$. The domain walls in this category host two chiral DMMs, transporting in opposite directions of the two chiral DMMs at the right and left outer edges (see Fig. 3b and Fig. S3a in SM [27]). The location of DMMs is determined by calculating their probability density $|\psi_R|^2$ in real-space and the transport direction is identified by their Fermi velocity. The third category has the interface between two FE TSC domains with the same $C$ and opposite sign of $\lambda$, which is characterized with chiral DMMs transporting oppositely at two outer edges and helical DMMs at the inner FE-FE domain wall (see Fig. 3c and Fig. S3b in SM [27]). The helical DMMs is similar to the ones around the interface where an s-wave pairing gap receives a π-phase shift on the surface of TI [6], but formed by interfacing two TSC domains of same Chern number but with opposite FE polarization that induces a π-phase difference in their pairing gap. The last category is interfaced by two FE TSC domains with opposite sign of $C$ and same sign of OPP $\lambda$. The transport of their domain-wall DMMs looks like that of the second category except the wave functions of two DMMs at the FM-FM domain wall hybridize (see Fig. 3d and Fig. S3c in SM [27]), forming fused DMMs. Clearly, in addition to FM, FE polarization provides another maneuverable knob to tune the transport of DMMs for FE TSC, which is robust against fluctuations around the domain wall (see Note S1 of SM [27]).

## C. First-principles calculation of α-In$_2$Se$_3$-based FE TSC

FEs are well-known functional materials with spontaneous FE polarization switchable by external electrical field. Recently, 2D FE materials have received increasing attention and been extensively explored in layered van der Waals materials [34-36], including the experimentally verified CuInP$_2$S$_6$ [37], SnTe [38], SnS [39], α-In$_2$Se$_3$ [40-42], β-In$_2$Se$_3$ [43], β′-In$_2$Se$_3$ [44], $d$1T-MoTe$_2$ [45], 1T″-MoS$_2$ layers [46], sliding 1T″-ReS$_2$ multilayers [47], 1T′-WTe$_2$ trilayer [48], $T_d$-WTe$_2$ bilayer [49], 1H

TMD multi-layers [50], few-layer InSe [51], stacked bilayer boron nitride (BN) [52,53], BN/Bernal-stacked bilayer graphene/BN [54], *etc*. Here we focus on demonstrating an $\alpha$-In$_2$Se$_3$-based FE TSC, as an example to implement our theoretical prediction.

The $\alpha$-In$_2$Se$_3$ monolayer consists of five atomic planes (Se-In-Se-In-Se), where the shifting of central Se plane reverses the out-of-plane FE polarization (Fig. S8a in SM [27]). First-principles calculations (see Note S2 of SM [27]) show the polarization leads to spin-splitting of bands (Fig. S8b in SM [27]) and opposite FE polarization can be distinguished from the distribution of spin expectation values in momentum space. Taking the lowest conduction band as an example, one can see the chirality of spin textures is opposite for the upward (Fig. S8c in SM [27]) and downward (Fig. S8d in SM [27]) FE polarization. This switchable chirality is what simulated in our TB model by changing the sign of $\lambda$.

Since electrons can be doped by gating [55] or built-in depolarization field [56] in $\alpha$-In$_2$Se$_3$, we consider the *s*-wave pairing condensed in the conduction band of $\alpha$-In$_2$Se$_3$ near the $\Gamma$ point upon the proximity effect with a superconductor substrate. We concretely demonstrate different TSC domains by considering a nanoribbon made of $\alpha$-In$_2$Se$_3$ with opposite FE polarizations at right and left domains (Fig. 4a), while a magnetic field can be induced by interfacing with a FM layer [57-61]. By constructing a first-principles BdG Hamiltonian based on Wannier functions (see Note S3 of SM [27]), we successfully reproduce the DMMs at the FE/FM-FE/FM (Fig. 4b) and FE-FE (Fig. 4c) domain walls, whose transport features and distributions are in excellent agreement with the TB model calculations (Fig. 3b and 3c).

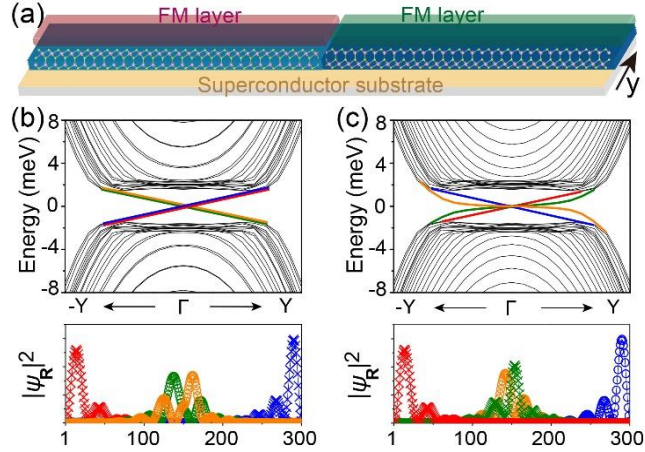

FIG. 4. Characterization of $\alpha$-In$_2$Se$_3$-based FE TSC. (a) Schematic diagram of $\alpha$-In$_2$Se$_3$-based FE TSC nanoribbon with opposite FE polarizations at the left and right domains. (b-c) Nanoribbon BdG quasi-particle band structure (upper panel) and real-space DMM distribution $|\psi_R|^2$ (lower panel) calculated for the nanoribbon with (b) opposite and (c) uniformed FM fields at left and right domains. The DMMs inside the pairing gap (upper panels) and the corresponding real-space distributions (lower panels) are plotted using the same color. $|\psi_R|^2$ peaks around the outer edges (position 1 or 300) and around the inner domain wall (position 150). Symbols of crosses and circles indicate the DMMs transporting in the +*y*- and −*y*-direction, respectively.

## III. DISCUSSION AND PERSPECTIVE

In general, FE TSC requires a FE metallic state with preferably large SOC. There are recent progress made in 2D FE metals [62-64], and gating may be applied to dope the FE semiconductor into metallic state, such as $\alpha$-$In_2Se_3$ [55,56], bilayer $T_d$-$WTe_2$ [49], and multi-layer $WSe_2$ or $MoS_2$ [50]. Another requirement is introducing SC into the FE layer via proximity effect, which has been long shown experimentally [65,66]. In addition, experiments have made FE/SC heterostructures with FE layer modulating SC in bilayer $T_d$-$MoTe_2$ [67], $Pb(Zr_xTi_{1-x})O_3$/$GdBa_2Cu_3O_{7-x}$ [68], Nb-doped-$SrTiO_3$/$Pb(Zr,Ti)O_3$ [69], and $BiFeO_3$/$YBa_2Cu_3O_{7-\delta}$ [70], which indirectly support our proposal. The last requirement is the FM field $V_z$, with tunable magnitude and direction, which is expected to be acquired by coupling FEs with FMs via heterostructures formed by vastly available FE and FM candidates [57-61]. Moreover, we expect that stoichiometric 2D multiferroics is also plausible to trigger the FE TSC phase. Given the experimental realization of multiferroicity in $CuCrP_2S_6$ [71], distorted monolayer $ReS_2$ [72], iron-doped $\alpha$-$In_2Se_3$ [73], single atomic layer of $NiI_2$ [74], atomically thin $\varepsilon$-$Fe_2O_3$ [75], and oxide-based 2D electron gas [76], we suggest future efforts to be made along this direction.

Because intrinsic TSC is very rare, much recent attention has been paid to extrinsic TSC employing proximity effect [8-12], While the existence of TSC and DMM has been experimentally confirmed [23-26], our proposed FE TSC paves the way to tuning the location, chirality, coupling, and motion of DMMs by using electrical field, which should stimulate future efforts along this interesting and exciting direction. We point out that the FE TSC phases are not limited to OPP gap of $\pm k_\pm$ we discuss here, because the polar point groups of FEs induce not only Rashba SOC, but also other forms of antisymmetric SOC and their mixtures [77]. This may further expand the diversity of OPP gaps, and hence the FE TSC domains. We also mention that the domain walls are actively explored in the field of "domain-wall nanoelectronics" [78] for potential applications of low-energy electronics in memory, logic, and brain-inspired neuromorphic computing. Here we propose domain-wall DMMs in FE TSC which may open a door to domain-wall quantum computation.

**Acknowledgements**

X.Z. and P.Z. acknowledge the financial support by the National Natural Science Foundation of China (No. 12004357) and the Natural Science Foundation of Shandong Province (No. ZR2020QA053). F.L. acknowledges financial support from DOE-BES (No. DE-FG02-04ER46148).

## References

[1] X.-L. Qi and S.-C. Zhang, Topological insulators and superconductors, Rev. Mod. Phys. **83**, 1057 (2011).

[2] M. Sato and Y. Ando, Topological superconductors: a review, Rep. Prog. Phys. **80**,

076501 (2017).

[3] Y. Hu and C. L. Kane, Fibonacci Topological Superconductor, Phys. Rev. Lett. **120**, 066801 (2018).

[4] B. Lian, X.-Q. Sun, A. Vaezi, X.-L. Qi, and S.-C. Zhang, Topological quantum computation based on chiral Majorana fermions, Proc. Natl. Acad. Sci. U.S.A. **115**, 10938 (2018).

[5] C. W. J. Beenakker, P. Baireuther, Y. Herasymenko, I. Adagideli, L. Wang, and A. R. Akhmerov, Deterministic Creation and Braiding of Chiral Edge Vortices, Phys. Rev. Lett. **122**, 146803 (2019).

[6] N. Read and D. Green, Paired states of fermions in two dimensions with breaking of parity and time-reversal symmetries and the fractional quantum Hall effect, Phys. Rev. B **61**, 10267 (2000).

[7] A. Y. Kitaev, Unpaired Majorana fermions in quantum wires, Phys.-Usp+ **44**, 131 (2001).

[8] L. Fu and C. L. Kane, Superconducting Proximity Effect and Majorana Fermions at the Surface of a Topological Insulator, Phys. Rev. Lett. **100**, 096407 (2008).

[9] X.-L. Qi, T. L. Hughes, and S.-C. Zhang, Chiral topological superconductor from the quantum Hall state, Phys. Rev. B **82**, 184516 (2010).

[10] J. Wang, Q. Zhou, B. Lian, and S.-C. Zhang, Chiral topological superconductor and half-integer conductance plateau from quantum anomalous Hall plateau transition, Phys. Rev. B **92**, 064520 (2015).

[11] M. Sato, Y. Takahashi, and S. Fujimoto, Non-Abelian Topological Order in *s*-Wave Superfluids of Ultracold Fermionic Atoms, Phys. Rev. Lett. **103**, 020401 (2009).

[12] J. D. Sau, R. M. Lutchyn, S. Tewari, and S. Das Sarma, Generic New Platform for Topological Quantum Computation Using Semiconductor Heterostructures, Phys. Rev. Lett. **104**, 040502 (2010).

[13] J. Röntynen and T. Ojanen, Topological Superconductivity and High Chern Numbers in 2D Ferromagnetic Shiba Lattices, Phys. Rev. Lett. **114**, 236803 (2015).

[14] J. Li, T. Neupert, Z. Wang, A. H. MacDonald, A. Yazdani, and B. A. Bernevig, Two-dimensional chiral topological superconductivity in Shiba lattices, Nat. Commun. **7**, 12297 (2016).

[15] A. Fornieri, A. M. Whiticar, F. Setiawan, E. Portolés, A. C. C. Drachmann, A. Keselman, S. Gronin, C. Thomas, T. Wang, R. Kallaher, G. C. Gardner, E. Berg, M. J. Manfra, A. Stern, C. M. Marcus, and F. Nichele, Evidence of topological superconductivity in planar Josephson junctions, Nature **569**, 89 (2019).

[16] H. Ren, F. Pientka, S. Hart, A. T. Pierce, M. Kosowsky, L. Lunczer, R. Schlereth, B. Scharf, E. M. Hankiewicz, L. W. Molenkamp, B. I. Halperin, and A. Yacoby, Topological superconductivity in a phase-controlled Josephson junction, Nature **569**, 93 (2019).

[17] S. Kawabata, Y. Asano, Y. Tanaka, A. A. Golubov, and S. Kashiwaya, Josephson $\pi$ State in a Ferromagnetic Insulator, Phys. Rev. Lett. **104**, 117002 (2010).

[18] C. Schrade, A. A. Zyuzin, J. Klinovaja, and D. Loss, Proximity-Induced $\pi$ Josephson Junctions in Topological Insulators and Kramers Pairs of Majorana Fermions, Phys. Rev. Lett. **115**, 237001 (2015).

[19] Z. Wang, J. O. Rodriguez, L. Jiao, S. Howard, M. Graham, G. D. Gu, T. L. Hughes, D. K. Morr, and V. Madhavan, Evidence for dispersing 1D Majorana channels in an iron-based superconductor, Science **367**, 104 (2020).

[20] R. Song, P. Zhang, and N. Hao, Phase-Manipulation-Induced Majorana Mode and Braiding Realization in Iron-Based Superconductor Fe(Te,Se), Phys. Rev. Lett. **128**, 016402 (2022).

[21] Y. Huang and C.-K. Chiu, Helical Majorana edge mode in a superconducting antiferromagnetic quantum spin Hall insulator, Physical Review B **98**, 081412 (2018).

[22] X. Zhang and F. Liu, Prediction of Majorana edge states from magnetized topological surface states, Phys. Rev. B **103**, 024405 (2021).

[23] J. Shen, J. Lyu, J. Z. Gao, Y.-M. Xie, C.-Z. Chen, C.-w. Cho, O. Atanov, Z. Chen, K. Liu, Y. J. Hu, K. Y. Yip, S. K. Goh, Q. L. He, L. Pan, K. L. Wang, K. T. Law, and R. Lortz, Spectroscopic fingerprint of chiral Majorana modes at the edge of a quantum anomalous Hall insulator/superconductor heterostructure, Proc. Natl. Acad. Sci. U.S.A. **117**, 238 (2020).

[24] G. C. Ménard, S. Guissart, C. Brun, R. T. Leriche, M. Trif, F. Debontridder, D. Demaille, D. Roditchev, P. Simon, and T. Cren, Two-dimensional topological superconductivity in Pb/Co/Si(111), Nat. Commun. **8**, 2040 (2017).

[25] A. Palacio-Morales, E. Mascot, S. Cocklin, H. Kim, S. Rachel, D. K. Morr, and R. Wiesendanger, Atomic-scale interface engineering of Majorana edge modes in a 2D magnet-superconductor hybrid system, Sci. Adv. **5**, eaav6600.

[26] S. Kezilebieke, M. N. Huda, V. Vaňo, M. Aapro, S. C. Ganguli, O. J. Silveira, S. Głodzik, A. S. Foster, T. Ojanen, and P. Liljeroth, Topological superconductivity in a van der Waals heterostructure, Nature **588**, 424 (2020).

[27] See Supplemental Material at http://link.aps.org/ for further discussions on the robustness of DMMs tuned by FM and/or FE polarization, details of first-principles calculations on $\alpha$-In$_2$Se$_3$-based FE TSC, and supplemental figures.

[28] P. Hohenberg and W. Kohn, Inhomogeneous Electron Gas, Phys. Rev. **136**, B864 (1964).

[29] W. Kohn and L. J. Sham, Self-Consistent Equations Including Exchange and Correlation Effects, Phys. Rev. **140**, A1133 (1965).

[30] J. P. Perdew, K. Burke, and M. Ernzerhof, Generalized Gradient Approximation Made Simple, Phys. Rev. Lett. **77**, 3865 (1996).

[31] P. E. Blöchl, Projector augmented-wave method, Phys. Rev. B **50**, 17953 (1994).

[32] A. A. Mostofi, J. R. Yates, G. Pizzi, Y.-S. Lee, I. Souza, D. Vanderbilt, and N. Marzari, An updated version of wannier90: A tool for obtaining maximally-localised Wannier functions, Comput. Phys. Commun. **185**, 2309 (2014).

[33] X. Zhang, K.-H. Jin, J. Mao, M. Zhao, Z. Liu, and F. Liu, Prediction of intrinsic topological superconductivity in Mn-doped GeTe monolayer from first-principles, npj Comput. Mater. **7**, 44 (2021).

[34] D. Zhang, P. Schoenherr, P. Sharma, and J. Seidel, Ferroelectric order in van der Waals layered materials, Nat. Rev. Mater. **8**, 25 (2023).

[35] M. Wu and J. Li, Sliding ferroelectricity in 2D van der Waals materials: Related physics and future opportunities, Proc. Natl. Acad. Sci. U.S.A. **118**, e2115703118

(2021).

[36] L. Qi, S. Ruan, and Y.-J. Zeng, Review on Recent Developments in 2D Ferroelectrics: Theories and Applications, Adv. Mater. **33**, 2005098 (2021).

[37] A. Belianinov, Q. He, A. Dziaugys, P. Maksymovych, E. Eliseev, A. Borisevich, A. Morozovska, J. Banys, Y. Vysochanskii, and S. V. Kalinin, $CuInP_2S_6$ Room Temperature Layered Ferroelectric, Nano Lett. **15**, 3808 (2015).

[38] K. Chang, J. Liu, H. Lin, N. Wang, K. Zhao, A. Zhang, F. Jin, Y. Zhong, X. Hu, W. Duan, Q. Zhang, L. Fu, Q.-K. Xue, X. Chen, and S.-H. Ji, Discovery of robust in-plane ferroelectricity in atomic-thick SnTe, Science **353**, 274 (2016).

[39] N. Higashitarumizu, H. Kawamoto, C.-J. Lee, B.-H. Lin, F.-H. Chu, I. Yonemori, T. Nishimura, K. Wakabayashi, W.-H. Chang, and K. Nagashio, Purely in-plane ferroelectricity in monolayer SnS at room temperature, Nat. Commun. **11**, 2428 (2020).

[40] Y. Zhou, D. Wu, Y. Zhu, Y. Cho, Q. He, X. Yang, K. Herrera, Z. Chu, Y. Han, M. C. Downer, H. Peng, and K. Lai, Out-of-Plane Piezoelectricity and Ferroelectricity in Layered $\alpha$-$In_2Se_3$ Nanoflakes, Nano Lett. **17**, 5508 (2017).

[41] C. Cui, W.-J. Hu, X. Yan, C. Addiego, W. Gao, Y. Wang, Z. Wang, L. Li, Y. Cheng, P. Li, X. Zhang, H. N. Alshareef, T. Wu, W. Zhu, X. Pan, and L.-J. Li, Intercorrelated In-Plane and Out-of-Plane Ferroelectricity in Ultrathin Two-Dimensional Layered Semiconductor $In_2Se_3$, Nano Lett. **18**, 1253 (2018).

[42] J. Xiao, H. Zhu, Y. Wang, W. Feng, Y. Hu, A. Dasgupta, Y. Han, Y. Wang, D. A. Muller, L. W. Martin, P. Hu, and X. Zhang, Intrinsic Two-Dimensional Ferroelectricity with Dipole Locking, Phys. Rev. Lett. **120**, 227601 (2018).

[43] Z. Zhang, J. Nie, Z. Zhang, Y. Yuan, Y.-S. Fu, and W. Zhang, Atomic Visualization and Switching of Ferroelectric Order in $\beta$-$In_2Se_3$ Films at the Single Layer Limit, Adv. Mater. **34**, 2106951 (2022).

[44] C. Zheng, L. Yu, L. Zhu, J. L. Collins, D. Kim, Y. Lou, C. Xu, M. Li, Z. Wei, Y. Zhang, M. T. Edmonds, S. Li, J. Seidel, Y. Zhu, J. Z. Liu, W.-X. Tang, and M. S. Fuhrer, Room temperature in-plane ferroelectricity in van der Waals $In_2Se_3$, Sci. Adv. **4**, eaar7720.

[45] S. Yuan, X. Luo, H. L. Chan, C. Xiao, Y. Dai, M. Xie, and J. Hao, Room-temperature ferroelectricity in $MoTe_2$ down to the atomic monolayer limit, Nat. Commun. **10**, 1775 (2019).

[46] A. Lipatov, P. Chaudhary, Z. Guan, H. Lu, G. Li, O. Crégut, K. D. Dorkenoo, R. Proksch, S. Cherifi-Hertel, D.-F. Shao, E. Y. Tsymbal, J. Íñiguez, A. Sinitskii, and A. Gruverman, Direct observation of ferroelectricity in two-dimensional $MoS_2$, npj 2D Mater. Appl. **6**, 18 (2022).

[47] Y. Wan, T. Hu, X. Mao, J. Fu, K. Yuan, Y. Song, X. Gan, X. Xu, M. Xue, X. Cheng, C. Huang, J. Yang, L. Dai, H. Zeng, and E. Kan, Room-Temperature Ferroelectricity in 1T'-$ReS_2$ Multilayers, Phys. Rev. Lett. **128**, 067601 (2022).

[48] Z. Fei, W. Zhao, T. A. Palomaki, B. Sun, M. K. Miller, Z. Zhao, J. Yan, X. Xu, and D. H. Cobden, Ferroelectric switching of a two-dimensional metal, Nature **560**, 336 (2018).

[49] S. C. de la Barrera, Q. Cao, Y. Gao, Y. Gao, V. S. Bheemarasetty, J. Yan, D. G. Mandrus, W. Zhu, D. Xiao, and B. M. Hunt, Direct measurement of ferroelectric

polarization in a tunable semimetal, Nat. Commun. **12**, 5298 (2021).

[50] S. Deb, W. Cao, N. Raab, K. Watanabe, T. Taniguchi, M. Goldstein, L. Kronik, M. Urbakh, O. Hod, and M. Ben Shalom, Cumulative polarization in conductive interfacial ferroelectrics, Nature **612**, 465 (2022).

[51] H. Hu, H. Wang, Y. Sun, J. Li, J. Wei, D. Xie, and H. Zhu, Out-of-plane and in-plane ferroelectricity of atom-thick two-dimensional InSe, Nanotechnology **32**, 385202 (2021).

[52] K. Yasuda, X. Wang, K. Watanabe, T. Taniguchi, and P. Jarillo-Herrero, Stacking-engineered ferroelectricity in bilayer boron nitride, Science **372**, 1458 (2021).

[53] M. Vizner Stern, Y. Waschitz, W. Cao, I. Nevo, K. Watanabe, T. Taniguchi, E. Sela, M. Urbakh, O. Hod, and M. Ben Shalom, Interfacial ferroelectricity by van der Waals sliding, Science **372**, 1462 (2021).

[54] Z. Zheng, Q. Ma, Z. Bi, S. de la Barrera, M.-H. Liu, N. Mao, Y. Zhang, N. Kiper, K. Watanabe, T. Taniguchi, J. Kong, W. A. Tisdale, R. Ashoori, N. Gedik, L. Fu, S.-Y. Xu, and P. Jarillo-Herrero, Unconventional ferroelectricity in moiré heterostructures, Nature **588**, 71 (2020).

[55] J. R. Rodriguez, W. Murray, K. Fujisawa, S. H. Lee, A. L. Kotrick, Y. Chen, N. McKee, S. Lee, M. Terrones, S. Trolier-McKinstry, T. N. Jackson, Z. Mao, Z. Liu, and Y. Liu, Electric field induced metallic behavior in thin crystals of ferroelectric $\alpha$-$In_2Se_3$, Appl. Phys. Lett. **117**, 052901 (2020).

[56] C. Ke, J. Huang, and S. Liu, Two-dimensional ferroelectric metal for electrocatalysis, Mater. Horizons **8**, 3387 (2021).

[57] W. Sun, W. Wang, D. Chen, Z. Cheng, and Y. Wang, Valence mediated tunable magnetism and electronic properties by ferroelectric polarization switching in 2D $FeI_2/In_2Se_3$ van der Waals heterostructures, Nanoscale **11**, 9931 (2019).

[58] C. Gong, E. M. Kim, Y. Wang, G. Lee, and X. Zhang, Multiferroicity in atomic van der Waals heterostructures, Nat. Commun. **10**, 2657 (2019).

[59] F. Xue, Z. Wang, Y. Hou, L. Gu, and R. Wu, Control of magnetic properties of $MnBi_2Te_4$ using a van der Waals ferroelectric $III_2$-$VI_3$ film and biaxial strain, Phys. Rev. B **101**, 184426 (2020).

[60] B. Yang, B. Shao, J. Wang, Y. Li, C. Yam, S. Zhang, and B. Huang, Realization of semiconducting layered multiferroic heterojunctions via asymmetrical magnetoelectric coupling, Phys. Rev. B **103**, L201405 (2021).

[61] K. Dou, W. Du, Y. Dai, B. Huang, and Y. Ma, Two-dimensional magnetoelectric multiferroics in a $MnSTe/In_2Se_3$ heterobilayer with ferroelectrically controllable skyrmions, Phys. Rev. B **105**, 205427 (2022).

[62] W. X. Zhou and A. Ariando, Review on ferroelectric/polar metals, Jpn J. Appl. Phys. **59**, SI0802 (2020).

[63] X.-Y. Ma, H.-Y. Lyu, K.-R. Hao, Y.-M. Zhao, X. Qian, Q.-B. Yan, and G. Su, Large family of two-dimensional ferroelectric metals discovered via machine learning, Sci. Bull. **66**, 233 (2021).

[64] D. Hickox-Young, D. Puggioni, and J. M. Rondinelli, Polar metals taxonomy for materials classification and discovery, Phys. Rev. Mater. **7**, 010301 (2023).

[65] C. Huang, B. T. Zhou, H. Zhang, B. Yang, R. Liu, H. Wang, Y. Wan, K. Huang, Z.

Liao, E. Zhang, S. Liu, Q. Deng, Y. Chen, X. Han, J. Zou, X. Lin, Z. Han, Y. Wang, K. T. Law, and F. Xiu, Proximity-induced surface superconductivity in Dirac semimetal $Cd_3As_2$, Nat. Commun. **10**, 2217 (2019).

[66] C. Li, Y.-F. Zhao, A. Vera, O. Lesser, H. Yi, S. Kumari, Z. Yan, C. Dong, T. Bowen, K. Wang, H. Wang, J. L. Thompson, K. Watanabe, T. Taniguchi, D. Reifsnyder Hickey, Y. Oreg, J. A. Robinson, C.-Z. Chang, and J. Zhu, Proximity-induced superconductivity in epitaxial topological insulator/graphene/gallium heterostructures, Nat. Mater. **22**, 570 (2023).

[67] A. Jindal, A. Saha, Z. Li, T. Taniguchi, K. Watanabe, J. C. Hone, T. Birol, R. M. Fernandes, C. R. Dean, A. N. Pasupathy, and D. A. Rhodes, Coupled ferroelectricity and superconductivity in bilayer $T_d$-$MoTe_2$, Nature **613**, 48 (2023).

[68] C. H. Ahn, S. Gariglio, P. Paruch, T. Tybell, L. Antognazza, and J. M. Triscone, Electrostatic Modulation of Superconductivity in Ultrathin $GdBa_2Cu_3O_{7-x}$ Films, Science **284**, 1152 (1999).

[69] K. S. Takahashi, M. Gabay, D. Jaccard, K. Shibuya, T. Ohnishi, M. Lippmaa, and J. M. Triscone, Local switching of two-dimensional superconductivity using the ferroelectric field effect, Nature **441**, 195 (2006).

[70] A. Crassous, R. Bernard, S. Fusil, K. Bouzehouane, D. Le Bourdais, S. Enouz-Vedrenne, J. Briatico, M. Bibes, A. Barthélémy, and J. E. Villegas, Nanoscale Electrostatic Manipulation of Magnetic Flux Quanta in Ferroelectric/Superconductor $BiFeO_3$/$YBa_2Cu_3O_{7-\delta}$ Heterostructures, Phys. Rev. Lett. **107**, 247002 (2011).

[71] Y. Lai, Z. Song, Y. Wan, M. Xue, C. Wang, Y. Ye, L. Dai, Z. Zhang, W. Yang, H. Du, and J. Yang, Two-dimensional ferromagnetism and driven ferroelectricity in van der Waals $CuCrP_2S_6$, Nanoscale **11**, 5163 (2019).

[72] J. Zhang, S. Wu, Y. Shan, J. Guo, S. Yan, S. Xiao, C. Yang, J. Shen, J. Chen, L. Liu, and X. Wu, Distorted Monolayer $ReS_2$ with Low-Magnetic-Field Controlled Magnetoelectricity, ACS Nano **13**, 2334 (2019).

[73] H. Yang, L. Pan, M. Xiao, J. Fang, Y. Cui, and Z. Wei, Iron-doping induced multiferroic in two-dimensional $In_2Se_3$, Sci. China Mater. **63**, 421 (2020).

[74] Q. Song, C. A. Occhialini, E. Ergeçen, B. Ilyas, D. Amoroso, P. Barone, J. Kapeghian, K. Watanabe, T. Taniguchi, A. S. Botana, S. Picozzi, N. Gedik, and R. Comin, Evidence for a single-layer van der Waals multiferroic, Nature **602**, 601 (2022).

[75] Y. Wang, P. Wang, H. Wang, B. Xu, H. Li, M. Cheng, W. Feng, R. Du, L. Song, X. Wen, X. Li, J. Yang, Y. Cai, J. He, Z. Wang, and J. Shi, Room-Temperature Magnetoelectric Coupling in Atomically Thin $\varepsilon$-$Fe_2O_3$, Adv. Mater. **35**, 2209465 (2023).

[76] J. Bréhin, Y. Chen, M. D'Antuono, S. Varotto, D. Stornaiuolo, C. Piamonteze, J. Varignon, M. Salluzzo, and M. Bibes, Coexistence and coupling of ferroelectricity and magnetism in an oxide two-dimensional electron gas, Nat. Phys. **19**, 823 (2023).

[77] X. Zhang, J. Liu, and F. Liu, Topological Superconductivity Based on Antisymmetric Spin–Orbit Coupling, Nano Lett. **22**, 9000 (2022).

[78] P. Sharma, T. S. Moise, L. Colombo, and J. Seidel, Roadmap for Ferroelectric Domain Wall Nanoelectronics, Adv. Funct. Mater. **32**, 2110263 (2022).